\documentclass[twocolumn,showpacs,prb,amsmath,amssymb,floatfix]{revtex4}
\usepackage{dcolumn}
\usepackage{amsmath,amssymb}
\usepackage{dsfont}
\usepackage{bm}
\usepackage{epsfig}

\newcommand{\nn}{\nonumber}
\newcommand{\bea}{\begin{eqnarray}}
\newcommand{\ena}{\end{eqnarray}}

\begin{document}

\title{Optical conductivity of graphene in the presence of random lattice deformations}

\author{A. Sinner$^1$, A. Sedrakyan$^{1,2}$, and K. Ziegler$^1$}

\affiliation{$^1$Institute for Physics, Universit\"at Augsburg, Universit\"atsstr. 1,
D-86159, Augsburg, Germany\\
$^2$Yerevan Physics Institute, Br. Alikhanian 2, Yerevan 36, Armenia}


\date{October 20, 2010}

\begin{abstract}
\noindent
We study the influence of lattice deformations on the optical conductivity of a two-dimensional 
electron gas. Lattice deformations are taken into account by introducing a non-abelian gauge 
field into the Eucledian action of two-dimensional Dirac electrons. This is in analogy to the introduction of the 
gravitation in the four-dimensional quantum field theory. We examine the effect of these deformations on the averaged optical conductivity.
Within the perturbative theory up to second order we show that corrections of the conductivity due to the deformations 
cancel each other exactly. We argue that these corrections vanish to any order in perturbative expansion.

\end{abstract}
\maketitle

\section{Introduction}

Graphene, a two-dimensional sheet of carbon atoms forming a honeycomb lattice,
has outstanding electronic properties~\cite{Novoselov2005,Novoselov2006,Katsnelson2006}.
This is due to the fact that there are two bands that touch each other at two Dirac nodes.
Moreover, the low-energy
quasiparticles of undoped graphene experience a linear dispersion around two
Dirac nodes. Transport properties, characterized by the longitudinal conductivity at the Dirac nodes,
are quite robust and do not vary much from sample to sample. Exactly at
the Dirac point a minimal conductivity has been observed in a number of experiments~\cite{Novoselov2005,zhang05,Novoselov2006}.
There are two important questions regarding this minimal conductivity: (I) is the value of the
minimal conductivity ``universal'' (i.e. independent of additional modifications of the
graphene sheet such as ripples or impurities) and (II) what is its actual value in units
of $e^2/h$?
A discrepancy between the calculated conductivity of Dirac fermions and the experimentally observed
minimal conductivity of graphene by a factor of roughly $1/\pi$ has been the subject of a substantial 
number of publications. The central idea is that either disorder~\cite{Fradkin1986,Lee1993,Ludwig1994,ziegler97}
or electron-electron interaction~\cite{Mishchenko2007,Sheehy2007,Herbut2008,deJuan2010} may affect the 
value of the minimal conductivity. Moreover, the value of minimal conductivity at low temperatures depends on the order of
varies limits (e.g. frequency $\omega\to0$ and temperature $T\to0$) and is related to the scaling property
$\sigma_{min}(\omega,T)=\sigma_{min}(\omega/T)$~\cite{ziegler07}. Below we will employ the zero-temperature 
formalism which suggests $T\to0$ and $\omega\to0$. This yields for the DC limit of the AC conductivity the value  $\pi/2$~\cite{Ludwig1994,Mishchenko2007,ziegler07}. 

An additional problem in terms of disorder is that it is not clear what role is played by different types
of disorder. Since disorder, depending on its type, may break different internal symmetries 
of the Dirac Hamiltonian, a classification according of the different types is crucial. On the other hand, the origin of disorder 
in graphene can be different. Besides impurities inside the graphene sheet and in the substrate,
the deformation of the lattice (e.g. ripples) might be the main source of disorder~\cite{Ishigamui2007,Meyer2007,Stolyarova2007}. 
In general, it is believed that surface corrugations~\cite{Khveshchenko2008,Herbut2008,Cortijo2009} 
may influence electronic transport properties of graphene. It is crucial to notice
that lattice deformations do not break the chiral symmetry at the Dirac point, in contrast to potential
disorder or a random gap caused by a random deposition of hydrogen~\cite{rgap}. Therefore, it is expected
that this type of disorder has a rather weak effect on transport properties~\cite{dos}. 
This is supported by calculations, where the lattice deformations are approximated by an uncorrelated random vector potential
in the Dirac Hamiltonian~\cite{VozKatsGuin2010}. This type of disorder has no effect on
the minimal conductivity~\cite{Fradkin1986}. More recently, however, a more general theory of lattice
deformations with long-range correlations revealed a dramatic increase of the minimal conductivity
for weak disorder~\cite{Cortijo2009}. In this paper we will study a similar model by an 
alternative approach to check whether or not this dramatic increase of the minimal conductivity can be reproduced. 

First we consider the deformation of the graphene sheet in three dimensions 
and show that in the continuum limit the dynamics of the electrons on the two-dimensional surface 
is defined by the so-called induced Dirac action presented in Ref.~[\onlinecite{Kavalov1986}].  
In our approach the internal deformations of the graphene sheet and the deformations perpendicular to the sheet
direction are unified into one schema, while in the approach developed in
papers~\cite{Cortijo2009,Cortijo2007a,Cortijo2007b,Juan2007,VozKatsGuin2010} there are separate 
internal 2D gravity and additional non-abelian gauge fields. The deformations of the sheet in three dimensions
by local $SO(3)$ rotations of the basic vectors in our approach carry the degrees of freedom of the additional gauge field.

Then we develop a replica-trick based field theory to take the random character of 
surfaces into account and to calculate the average optical conductivity by a perturbative 
expansion. Our main result is that the random lattice deformations do not affect the robust character of minimal 
conductivity, contrary to the result presented in Ref.~[\onlinecite{Cortijo2009}].

\section{The model}
\label{sec:model}

\begin{figure}[t]
\centerline{\includegraphics[width=65mm,angle=0,clip]{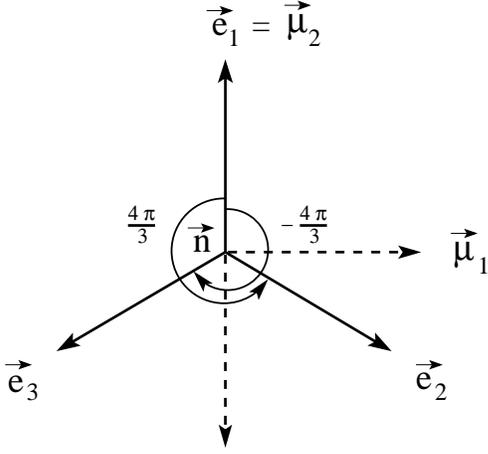}}
\caption{Hopping vectors on regular honeycomb lattice} 
\label{fig:0}
\end{figure}

\begin{figure}[t]
\centerline{\includegraphics[width=70mm,angle=0,clip]{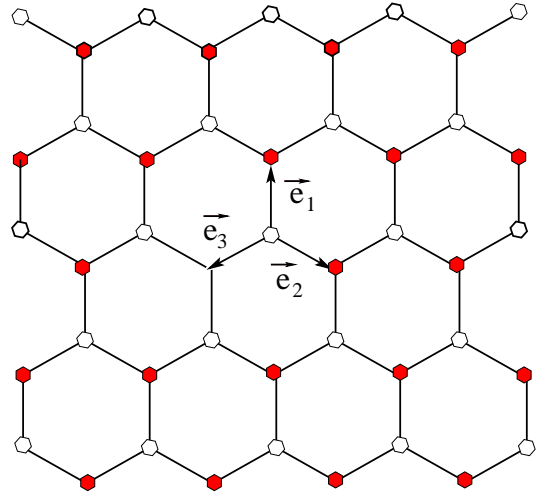}}
\caption{(Color online) Regular honeycomb lattice.} 
\label{fig:1}
\end{figure}

We depart from a model of hopping fermions on the regular 2D honeycomb lattice.
Honeycomb lattice has natural partition into two triangular sub-lattices and we mark electronic fields
associated with sites  of the sub-lattices as $(\bar{\psi}_{\vec{n},\alpha}, \psi_{\vec{n},\alpha}),\;\;\alpha=1,2$.
The action of electrons hopping on a line with the lattice spacing $|{\vec e}|$ reads
$$
{\cal S}[\bar\psi,\psi] = i\sum_{t,{\vec n}} ( \bar{\psi}^{}_{t,\vec{n}}\partial_t \psi^{}_{t,\vec{n}} + \bar{\psi}^{}_{t,\vec{n}}\gamma^{}_2 \psi^{}_{t,\vec{n}+{\vec e}}),
$$ 
but  when fermions change hopping direction in two dimensional space they fields should also be rotated by a corresponding 
angle (Fig.~\ref{fig:0}). On the honeycomb lattice (Fig.~\ref{fig:1}) we have
\begin{eqnarray}
\label{action1}
{\cal S}[{\bar \Psi},\Psi] = i\sum_{t,\vec{n},i} \left(\bar{\Psi}_{t,\vec{n}}\partial_t \Psi^{}_{t,\vec{n}} +
\bar{\Psi}_{t,\vec{n}}\gamma^{}_2 \Psi^{\prime}_{t,\vec{n}+\vec{e}_i}\right), \\
\Psi_{t,\vec{n}}=\left( \begin{array}{c}
\psi_{t,\vec{n},1}\\
\psi_{t,\vec{n},2}
\end{array}\right), \qquad i=1,2,3,\nn
\end{eqnarray}
where  $\gamma^{}_0,\gamma_j, \; j=1,2$ are Dirac matrices which are related to usual Pauli matrices via $\gamma^{}_1=\sigma^{}_2$, $\gamma^{}_2=\sigma^{}_1$, and $\gamma^{}_0=\sigma^{}_3$ and fields
\begin{subequations}
\bea
\label{psi1}
\Psi^{\prime}_{\vec{n}+\vec{e}_1}&=& \Psi_{\vec{n}+\vec{e}_1}=
e^{\vec{e}_1\cdot\vec{\partial}} \Psi_{\vec{n}}, \\
\Psi^{\prime}_{\vec{n}+\vec{e}_2}&=& e^{i \frac{2 \pi}{3} \gamma_0} \Psi_{\vec{n}+\vec{e}_2}=
e^{i \frac{2 \pi}{3} \gamma_0}e^{\vec{e}_2\cdot\vec{\partial}} \Psi_{\vec{n}}, \\
\Psi^{\prime}_{\vec{n}+\vec{e}_3}&=& e^{-i \frac{2 \pi}{3} \gamma_0} \Psi_{\vec{n}+\vec{e}_3}
= e^{-i \frac{2 \pi}{3} \gamma_0}e^{\vec{e}_3\cdot\vec{\partial}} \Psi_{\vec{n}},
\ena
\end{subequations}
are rotated by $\pm 4 \pi/3$ and translated by $ \vec{e}_{2,3}$ spinor representations of the rotation
group $SO(3)$.  In the paper by Semenoff~\cite{Semenoff1984} was shown that
the spectrum of low-energy excitations of the hopping fermions on
honeycomb lattice (corresponding to the continuum limit of the model)
coincides with the spectrum of Dirac fermions in 3D space. Below we will show that the continuum limit of the action of 
fermions hopping on honeycomb lattice Eq.~(\ref{action1}) is defined by the Dirac action in three dimensional coordinate space. 
This will allow to construct the continuum limit of the generalized hopping model on the randomly deformed lattice.

\begin{figure}[t]
\centerline{\includegraphics[width=70mm,angle=0,clip]{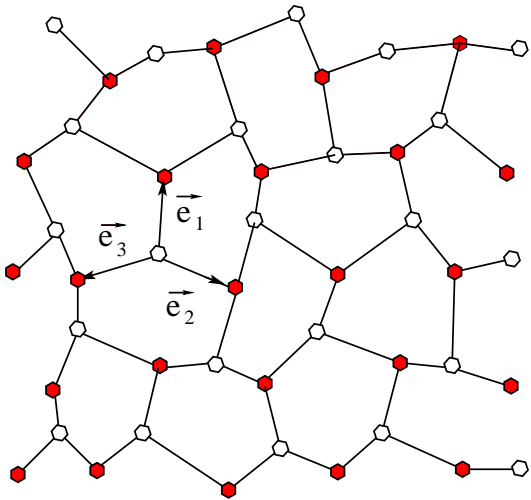}}
\caption{(Color online) Random honeycomb lattice.} 
\label{fig:2}
\end{figure}

In order to find a continuum limit of the action Eq.~(\ref{action1}) one expands translational operators
$e^{\vec{e}^{}_i\cdot\vec{\partial}}\simeq 1 + \vec{e}^{}_i\cdot\vec{\partial} $ and substitute Eqs.~(\ref{psi1})
for $\Psi$'s into the action Eq.~(\ref{action1}). Then after some simple algebra one will obtain:
\bea
\label{eq:action2}
&&{\cal S}[{\bar\Psi},\Psi] = i\sum_{t,\vec{n}} \bar{\Psi}_{t,\vec{n}} 
\left(
\partial^{}_t +\frac{3}{4} 
\gamma^i{\vec\mu}^{}_i\cdot
[\overleftarrow{\partial}-\overrightarrow{\partial}]
\right)
\Psi_{t,\vec{n}}\;\;\;\\
\label{eq:action2a}
&&\to
i\int d^2\xi dt~
\bar{\Psi}
\left(
\partial^{}_t +\frac{1}{2} 
\gamma^i{\vec\mu}^{}_i\cdot
[\overleftarrow{\partial}-\overrightarrow{\partial}]
\right)
\Psi,\;\;
\ena
where we introduced orthonormalized vectors  $\vec{\mu}^{}_1=(\vec{e}_2-\vec{e}_3)/\sqrt{3}$ and $\vec{\mu}^{}_2=\vec{e}_1$.
In the line (\ref{eq:action2a}) we have rescaled the fields and coordinates as $\Psi\to2/3\Psi$, $\xi\to3\xi/2$, $t\to t$ and $\mu^{}_i\to\mu^{}_i$. It is clear from Eq.~(\ref{eq:action2a}) that vectors $\vec{\mu}_a = \mu^i_a({\vec\xi})\hat{e}_i, \; a=1,2$, with $\hat{e}_i,\; i=1,2$ representing an orthonormal basis in the flat space play role of tetrads (vielbein) in a 2D plane with arbitrary coordinates $\xi_i, \; i=1,2$. Indeed, consider deformation of the honeycomb lattice (cf. Fig.~\ref{fig:2}) and attach to the sites a new coordinates $\xi_i^{\prime}$. Then the vectors $\mu^{i}_a,\; a=1,2,\; i=1,2$ will be connected with the same vectors in the old coordinate $\xi_i,\;i=1,2$ via 
\bea
\label{xi}
\mu_a^i({\vec\xi})=\frac{\partial \xi_j^{\prime}}{\partial \xi^{}_i}\mu_a^j({\vec\xi}^{\prime}).
\ena
We regard now the vectors $\mu_i^a$ as vielbeins in a 2D plane which obey the orthogonality relation $\mu_i^a\mu_{a,j}=\delta_{ij}$ and define the metric $\mu_i^a\mu_j^a=g_{ij}$. After integration by parts in Eq.~(\ref{eq:action2}) and using the relation
$\hat\mu_i \hat\mu_j=g_{ij}+\frac{i}{\sqrt{g}}\epsilon_{ij}\gamma^{}_0$ with 
$\hat{\mu}_i=\gamma^a \mu_i^a$ and $g={\rm det}[g_{ij}]$ 
one will obtain
\bea
\label{action3}
{\cal S}[{\bar\Psi},\Psi] = i\int d^2\xi dt~
\bar{\Psi}\left(\partial^{}_t +
\gamma^a \mu_a^{j}\big[\partial^{}_j-\frac{i}{2}\gamma^{}_0 \Gamma_j \big]\right)\Psi,\;\;\; 
\ena
where $\Gamma_j=\frac{i}{2 \sqrt{g}} \epsilon_{ab}\mu_a^k\nabla_j\mu_{k, b}$ is a standard spinor connection corresponding to the 
vielbein $\mu_a^j$ and $\nabla^{}_j$  denotes a covariant derivative. For a scalar function $f$ 
it reduces to a usual partial derivative: $\nabla^{}_i f= \partial^{}_i f$, while for a vector valued function $f^{}_j$ 
it is $\nabla^{}_i f^{}_j=\partial^{}_i f^{}_j + \Gamma^{k}_{ij} f_k$, where $\Gamma^{k}_{ij}$ represent 
Christoffel symbols. 

Let us now consider deformations of the honeycomb lattice in a three dimensional space~\cite{Kavalov1986, Sedrakyan1987}. This means
that two $\gamma_a,\; a=1,2$ matrices in a tangent plane become $\gamma$
matrices in a $SO(3)$ rotated plane which is tangent to curved surface at the point $\xi_i$:
\bea
\label{x}
\hat{x}_a(\xi_i)= U(\xi_i)^{-1}\gamma^{}_a U(\xi_i).
\ena
As it is shown in Ref.~[\onlinecite{Kavalov1986}] local rotations by $U({\vec\xi})$ produce a 2D surface embedded into 3D Euclidean space if
\bea
\label{rotation}
U^{-1}\partial^{}_\mu U= \frac{1}{4}\big(\hat{x}_a \partial^{}_\mu \hat{x}_a + \hat{n} \partial^{}_\mu \hat{n}\big),
\ena
where $\hat{x}_a=\mu_a^{\mu} \partial_\mu  x^{\alpha}\gamma^{}_{\alpha}$ and
$\hat{n}=n^{\alpha}\gamma^{}_{\alpha},\;\alpha=0,1,2$ being the tangent and normal to the surface 3D vectors
at the point $\xi$ respectively. This will occur since Eq.~(\ref{rotation}) will fulfill the Gauss-Codazzi
equations~\cite{Sedrakyan1987, Novikov1984}, which represents the necessary conditions for the surface $x^{\alpha}$ to be embedded into 3D Euclidean space.

Then we should rotate also fermionic fields by the same matrices $\Psi\rightarrow U \Psi$, after which the action becomes
\bea
\label{eq:action4}
{\cal S}
= i\int d^2{\xi}dt~
\bar{\Psi}U^{-1}\left(
\partial_t +
\gamma^a \mu_a^{\mu}
[\partial^{}_\mu-\frac{i}{2}\gamma^{}_0 \Gamma^{}_\mu]
\right)
U\Psi,\;\;
\ena
where $\Gamma_\mu=i {\rm Tr}(\hat{n}U^{-1}\partial_\mu U)$ 
is the spinor connection on the surface $x^{\alpha}$. By use of Eq.~(\ref{x}) this expression can be simplified essentially 
(see details in Refs.~[\onlinecite{Kavalov1986},\onlinecite{Sedrakyan1987}]) acquiring the form
\bea
\label{eq:reduced}
{\cal S}
[{\bar\Psi},\Psi] 
= i\int d^2{\xi}dt~
\bar{\Psi}
\left(
\partial^{}_t +\frac{1}{2}\sqrt{g} \hat{\gamma}^\mu
\big[\overleftarrow{\partial^{}_\mu}-\overrightarrow{\partial^{}_\mu}\big]
\right)
\Psi,\;\;\;
\ena
where 
\begin{equation}
\label{eq:Metric}
g^{}_{\nu\mu} = \partial^{}_\nu{\bf x}\cdot\partial^{}_\mu{\bf x} = \partial^{}_\nu x^{\alpha}\partial^{}_\mu x^{}_{\alpha}
\end{equation}
is the metric on the surface $x^{\alpha}({\vec\xi})$ 
induced by its embedding into 3D Euclidean space, $g$ denotes its determinant and 
\begin{equation}
\label{eq:xField} 
\hat{\gamma}^\mu = \partial^\mu{\bf x}\cdot{\bf\gamma}=\partial^\mu x^a_{}\gamma^{}_a
\end{equation}
represent the induced Dirac matrices~\cite{Kavalov1986}. In a flat space, i.e. for ${\bf x}({\vec\xi})={\bf x}_0+\hat{\bf e}^\mu\xi_\mu$ the induced metric reduces to a usual diagonal matrix. 
One can call action Eq.~(\ref{eq:reduced}) the induced Dirac action since the matrices $\hat{\gamma}^\mu$ are induced by embedding. The expression in Eq.~(\ref{eq:reduced}) is a generalization of 2D action Eq.~(\ref{eq:action2a}) to 3D space.

\section{Effective action for small corrugations}
\label{sec:effect}

Performing integration by parts in Eq.~(\ref{eq:reduced}) we arrive at 
\begin{eqnarray}
\label{eq:StartAction} 
{\cal S}[\bar\Psi,\Psi]
&=& i\int {d^2\xi dt}~\bar\Psi\left(\partial^{}_t+\sqrt{g}\hat{\gamma}^\mu[\partial^{}_\mu+\Gamma^{}_\mu]\right)\Psi.\;
\end{eqnarray}
Here, the quantity 
\begin{equation}
\label{eq:GammaVec}
\Gamma^{}_\mu = \frac{1}{2} \hat{\gamma}^\nu\nabla^{}_\mu\hat{\gamma}^{}_\nu
\end{equation}
plays the role of an induced spinor connection, where  
$\nabla^{}_\mu$ denotes the operator of covariant differentiation and is defined as~\cite{Kavalov1986, Sedrakyan1987}
$$
\nabla^{}_\mu(\cdots)=\frac{1}{\sqrt{g}}\partial^{}_\mu(\sqrt{g}\cdots).
$$

Let us derive the asymptotic action for small corrugations of the graphene sheet. In this case the surface ${\bf x}$ can be asymptotically represented as
\begin{equation}
\label{eq:xLinearizing} 
{\bf x}(\xi^{}_1,\xi^{}_2) \approx {\bf x}^{}_0 + \hat{\bf e}^\mu\xi^{}_\mu + {\bf x}^\prime(\xi^{}_1,\xi^{}_2),
\end{equation}
Plugging Eq.~(\ref{eq:xLinearizing}) into Eq.~(\ref{eq:Metric}) we obtain the asymptotics of the metric tensor:
\begin{eqnarray} 
\label{eq:MetricAsympt}
g^{}_{\nu\mu} \approx \delta^{}_{\nu\mu} + \epsilon^{}_{\nu\mu} + \epsilon^{}_{\mu\nu},
\end{eqnarray}
where 
\begin{eqnarray}
\epsilon^{}_{\nu\mu} = \hat{\bf e}^{}_\nu\cdot\partial^{}_\mu{\bf x}^\prime.
\end{eqnarray}
Thus the metric tensor is in general neither diagonal nor symmetric. Its determinant is found using common relations
\begin{equation}
\label{eq:MetricDeterminant} 
g \approx 1+2 \epsilon^{}_{11}+2 \epsilon^{}_{22} = 1+2 \epsilon^{}_{\nu\nu},
\end{equation}
and correspondingly its square root:
\begin{equation}
\label{eq:MetricDeterminantSqrt} 
\sqrt{g} \approx 1+\epsilon^{}_{\nu\nu}.
\end{equation}

Using Eqs.~(\ref{eq:xField}), (\ref{eq:xLinearizing} ) and (\ref{eq:MetricDeterminantSqrt}) we arrive at the effective action for small fluctuations $\epsilon$:
\begin{eqnarray}
\nonumber
{\cal S}^{}_0[\bar\Psi,\Psi;\epsilon] &\approx& i\intop{d^2\xi}{dt}~
\bar\Psi^{}\left(\gamma^{}_0\partial^{}_t + [1+\epsilon^{}_{\nu\nu}]\tilde\gamma^\mu\partial^{}_\mu \right)\Psi^{}\\
\label{eq:ActionFin} 
&+& \frac{i}{2}\intop{d^2\xi}{dt}~\bar\Psi^{}\tilde\gamma^\mu\partial^{}_\mu \epsilon^{}_{\nu\nu}\Psi^{},
\end{eqnarray}
with induced $\gamma-$matrices $\tilde\gamma^{\mu}=e^{\mu}_{a}\gamma^a$. 
For further purposes we associate the spatial fluctuations with a bosonic field
$$
\epsilon^{}_{\nu\nu} = \Lambda({\vec\xi},t),
$$
and its gradient with a static vector-disorder like term:
$$
\partial^{}_\mu \epsilon^{}_{\nu\nu} = \partial^{}_\mu\Lambda({\vec\xi},t) = B_\mu({\vec\xi},t).
$$
Hence the action formally becomes
\begin{eqnarray}
\nonumber
{\cal S}^{}_0[\bar\Psi,\Psi;\Lambda,{\bm B}] &\approx& i\intop{d^2\xi}{dt}~
\bar\Psi^{}\left(\gamma^{}_0\partial^{}_t + \tilde\gamma^\mu\partial^{}_\mu \right)\Psi^{}\\
\nonumber
&+& i\intop{d^2\xi}{dt}~\Lambda \bar\Psi^{}\tilde\gamma^\mu\partial^{}_\mu \Psi^{}\\
\label{eq:ActionFin2} 
&+& \frac{i}{2}\intop{d^2\xi}{dt}~\bar\Psi^{}\tilde\gamma^\mu B_\mu\Psi^{}.
\end{eqnarray}
The action derived this way reproduces the ansatz action considered in Ref.~[\onlinecite{Cortijo2009}]. 

Below we consider topologic defects in the flat space. Technically that means that we replace zweibeins $e^\mu_a$ by a unity-matrix. 
We are ultimately interested in the effect of this sort of the disorder on the optical conductivity. 
In order to perform such calculations we have to make some suggestions regarding the correlators of the introduced quantities. One  usually requires the vector disorder fields $B_\mu$ to be gaussian correlated, i.e. 
\begin{eqnarray}
\label{eq:AcorR1}
\langle B^{}_\mu({\vec\xi},t)\rangle &=& 0, \\
\label{eq:AcorR2}
\langle B^{}_\mu({\vec\xi},t)B^{}_\nu({\vec\xi}^\prime,t^\prime) \rangle &=& g^{2}_0\delta^{}_{\mu\nu}\delta(t)\delta(t^\prime)\delta({\vec\xi}-{\vec\xi}^\prime),
\end{eqnarray}
which guarantees that the vector associated with the random disorder is static. In Fourier-space these expressions read with the short-hand $Q=(q^{}_0,{\bm q})$:
\begin{eqnarray}
\label{eq:AcorF1}
\langle B^{}_\mu(Q)\rangle &=& 0, \\
\label{eq:AcorF2}
\langle B^{}_\mu(Q)B^{}_\nu(Q^\prime) \rangle &=& 
g^{2}_0(2\pi)^4\delta^{}_{\mu\nu}\delta(q^{}_0)\delta(Q^{}+Q^\prime). 
\end{eqnarray}
From Eqs.~(\ref{eq:AcorR2}) and (\ref{eq:AcorF2}) we are lead to the correlators of the scalars $\Lambda$, since we have an exact relationship
\begin{equation}
\langle B^{}_\mu(Q)B^{}_\nu(Q^\prime) \rangle = i^2 q^{}_\mu q^\prime_\nu \langle \Lambda(Q)\Lambda(Q^\prime)\rangle,
\end{equation}
which leads to 
\begin{equation}
\label{eq:contr}
\langle \Lambda(Q)\Lambda(Q^\prime)\rangle = \frac{g^{2}_0}{q^2+\mu^2}(2\pi)^4 \delta(q^{}_0)\delta(Q+Q^\prime),
\end{equation}
where we have introduced an infrared cutoff $\mu$ of the order of the inverse lattice spacing in order to avoid long wave-length divergences. Inverse Fourier transform yields for the $\langle\Lambda\Lambda\rangle-$correlator 
\begin{equation}
\label{eq:LLcorr}
\langle \Lambda({\vec\xi},t)\Lambda({\vec\xi}^\prime,t^\prime)\rangle = g^{2}_0 \delta(t)\delta(t^\prime)\log\left|\mu({\vec\xi}-{\vec\xi}^\prime)\right|.
\end{equation}
Furthermore we will always assume
\begin{equation}
\label{eq:Lcorr}
\langle \Lambda({\vec\xi},t)\rangle = \langle \Lambda(Q)\rangle = 0. 
\end{equation}
The Fourier transform of Eq.~(\ref{eq:ActionFin}) expressed in terms of the scalar fields $\Lambda$ only reads
\begin{eqnarray}
\nonumber
&&{\cal S}[\bar\Psi,\Psi,\Lambda] = -\intop_{Q}~\bar\Psi^{}_{Q}(q^{}_0\gamma^{}_0+\gamma\cdot{\bm q})\Psi^{}_{Q} \\
\label{eq:ActionFourier}
&&-  \intop_{Q} \intop_{P}~\Lambda^{}_{P}\bar\Psi^{}_{{P}+{Q}} \Gamma(P+Q,P,Q) \Psi^{}_{Q},
\end{eqnarray}
where $\gamma\cdot{\bm q}=\gamma^\mu q^{}_\mu$. The two-particle vertex is obtained from Eq.~(\ref{eq:ActionFourier}) 
in limit $\Lambda\to0$ by performing second order functional derivative with respect to the Grassmann fields:
\begin{equation}
\label{eq:TwoPoint}
\Gamma^{}(P,Q)=(2\pi)^3\delta\left(P-Q\right)G^{-1}_0(Q),
\end{equation}
where
\begin{eqnarray}
G^{-1}_0(Q) = q^{}_0\gamma^{}_0+\gamma\cdot{\bm q},
\end{eqnarray}
represents the inverse free propagator and correspondingly
\begin{equation}
\label{eq:FermiProp}
G^{}_0(Q)=\frac{q^{}_0\gamma^{}_0+\gamma\cdot{\bm q}}{q^2_0+q^2} 
\end{equation}
the free Dirac propagator. For the three-particle vertex function $\Gamma(K^{}_{\bar\Psi},K^{}_\Lambda,K^{}_\Psi)$ follows from Eq.~(\ref{eq:ActionFourier}) 
\begin{eqnarray}
\nonumber
\Gamma(K^{}_{\bar\Psi};K^{}_\Lambda,K^{}_\Psi) &=& (2\pi)^3\delta(K^{}_{\bar\Psi}-K^{}_{\Lambda}-K^{}_{\Psi})\\
\label{eq:Vertex}
&\times&\frac{1}{2}\gamma\cdot\left({\bm k}^{}_\Lambda+2{\bm k}^{}_{\Psi}\right).
\end{eqnarray}

Furthermore we have to augment Eq.~(\ref{eq:ActionFourier}) by the interaction between fermions and the radiation field
\begin{equation}
\label{eq:FermActInt}
{\cal S}^{}_{\rm opt}[\bar\Psi,\Psi,A] = -\intop_{P}\intop_{Q}~A^{}_P \bar\Psi^{}_{P+Q}\gamma^{}_0\Psi^{}_{Q},
\end{equation}
which suggests the presence of an electric field applied to the graphene sheet. Interaction Eq.~(\ref{eq:FermActInt}) gives rise to the optical conductivity due to polarization of the charge carriers. The corresponding bare vertex is defined as
\begin{equation}
\Gamma^{}_0(K^{}_{\bar\Psi},K^{}_A,K^{}_\Psi) = (2\pi)^3\delta(K^{}_{\bar\Psi}-K^{}_A-K^{}_\Psi)\gamma^{}_0.
\end{equation}

The full action acquires the form
\begin{equation}
\label{eq:CombActionFin}
\bar{\cal S}[\bar\Psi,\Psi,\Lambda,A] = {\cal S}[\bar\Psi,\Psi,\Lambda] + {\cal S}^{}_{\rm opt}[\bar\Psi,\Psi,A].
\end{equation}
The electron-gauge boson interaction renormalizes electronic spectrum and therefore should
have an effect on the response to the radiation field. 

\section{optical conductivity of graphene}

We consider first ideal graphene. The corresponding Eucledian action is obtained from Eq.~(\ref{eq:CombActionFin}) if we assume $\Lambda=0$:
\begin{eqnarray}
\nonumber
{\cal S}^{}_0[\bar\Psi,\Psi;{A}] = -\intop_{Q}~\bar\Psi^{}_{Q}(q^{}_0\gamma^{}_0+\gamma\cdot{\bm q})\Psi^{}_{Q}\\
\label{eq:Action} 
-\intop_{P}\intop_{Q}~A(P-Q) \bar\Psi^{}_{P}\gamma^{}_0\Psi^{}_{Q},
\end{eqnarray}

\begin{figure}[t]
\includegraphics[height=3.cm]{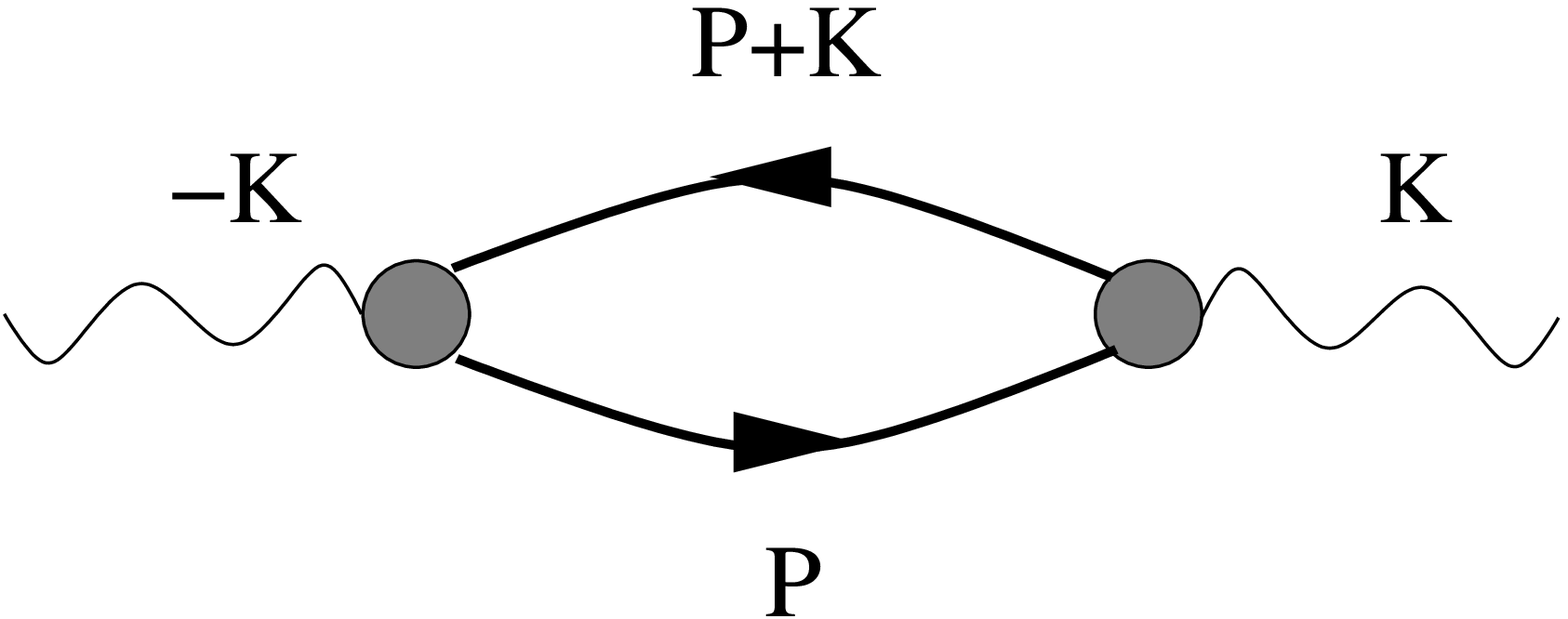}  
\caption{Bare polarization bubble} 
\label{fig:DDC}
\includegraphics[height=3.cm]{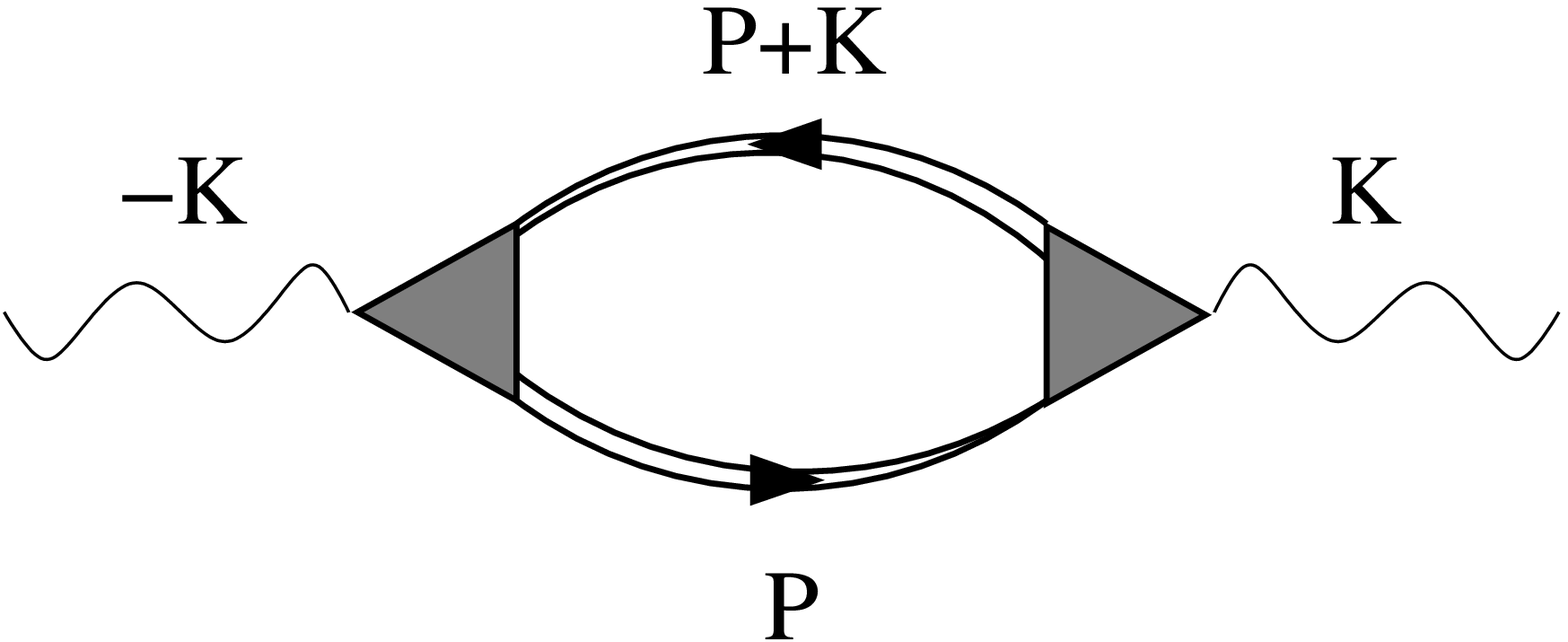}  
\caption{Dressed polarization bubble} 
\label{fig:DDCscreened}
\end{figure}

The optical conductivity of 2D Dirac electron gas can be calculated from the electronic polarization~\cite{Mishchenko2007,deJuan2010}
\begin{equation}
\label{eq:CondGen} 
\sigma^{}_0 = {4k^{}_0}\lim_{k\to0}\frac{1}{2}\frac{\partial^2}{\partial k^2}\Pi(K), 
\end{equation}
where $\Pi(K)$ denotes the irreducible polarization. Factor 4 in front of this expression arises from the taking both spin and valley degeneracy into account. To the leading order it is  given by the diagram shown in Fig.~\ref{fig:DDC}. Algebraically we have for the polarization bubble:
\begin{equation}
\label{eq:IrPol}
\Pi(K) = \intop_P~{\rm Tr}\left\{\gamma^{}_0 G_0(P)\gamma^{}_0 G^{}_0(K+P)\right\},
\end{equation}
with the bare Dirac propagators $G^{}_0(Q)$ defined in Eq.~(\ref{eq:FermiProp}). We give some details of the calculation in the Appendix. 
The irreducible polarization is obtained as
\begin{eqnarray}
\label{eq:IrPolRes}
\Pi(K) =\frac{1}{16}\frac{k^2}{\sqrt{k^2+k^2_0}},
\end{eqnarray}
and the optical conductivity in SI-units is
\begin{equation}
\label{eq:CondLine4} 
\sigma^{}_0 = \frac{1}{4} \frac{e^2}{\hbar} = \frac{\pi}{2} \frac{e^2}{h}. 
\end{equation}

In what follows we calculate corrections of the conductivity in Eq.~(\ref{eq:CondLine4}) due to lattice deformations described in Sec.~\ref{sec:effect}. To the leading order in momenta the renormalized inverse fermionic propagator can be written as
\begin{eqnarray}
\label{eq:DresProp}
G^{-1}(Q) = G^{-1}_0(Q) - \Sigma(Q) \approx Z^{-1}_{0} q^{}_0\gamma^{}_0 + Z^{-1}_{1}\gamma\cdot{\bm q},\;\;\;
\end{eqnarray}
with renormalization factors 
\begin{eqnarray}
\nonumber
Z^{-1}_{0,1} &=& 1 - \gamma^{}_{0,\mu}\left.\frac{\partial}{\partial q^{}_{0,\mu}}\Sigma(Q)\right|_{Q=0} \\
\label{eq:WaveFuncRen}
&=& \gamma^{}_{0,\mu}\left.\frac{\partial}{\partial q^{}_{0,\mu}}G^{-1}(Q)\right|_{Q=0},
\end{eqnarray}
where $\Sigma(Q)$ denotes the fermionic self-energy. On another hand, the dressed electron-photon vertex can be written in the following from:
\begin{equation}
\label{eq:DresVert}
\tilde\Gamma^{}_0(Q) \approx \tilde{e}\gamma^{}_0 + {\cal O}(Q),
\end{equation}
where $\tilde{e}$ denotes the renormalization of the elementary charge due to lattice deformations, such that the effective renormalized action reads
\begin{eqnarray}
\nonumber
\tilde{\cal S}^{}_0[\bar\Psi,\Psi;{A}] &\approx& -\intop_{Q}~\bar\Psi^{}_{Q}(Z^{-1}_0 q^{}_0\gamma^{}_0+Z^{-1}_1\gamma\cdot{\bm q})\Psi^{}_{Q}\\
\label{eq:ActionRen} 
&-&\tilde{e}\intop_{P}\intop_{Q}~A(P-Q) \bar\Psi^{}_{P}\gamma^{}_0\Psi^{}_{Q}.
\end{eqnarray}

The effect of the lattice defects on the optical conductivity can be calculated from the dressed polarization shown in Fig.~\ref{fig:DDCscreened}. The dressing effect of the lattice defects is taken into account by replacing  bare Greens functions $G^{}_0$ in Fig.~\ref{fig:DDC} by the full propagators $G$ defined in Eq.~(\ref{eq:DresProp}) and bare vertices $\gamma^{}_0$ by the dressed ones $\tilde\Gamma^{}_0$ from Eq.~(\ref{eq:DresVert}). Algebraically we obtain
\begin{eqnarray}
\nonumber
&&\tilde\Pi^{}(K) = \intop_P~{\rm Tr}\{\tilde\Gamma^{}_0 G(P) \tilde\Gamma^{}_0 G^{}(K+P)\}\\
&&=\frac{2\tilde{e}^2}{Z^{-2}_0}\intop_P~\frac{p^{}_0(p^{}_0+k^{}_0)-\alpha^2{\bm p}\cdot({\bm p}+{\bm k})}
{[p^2_0+\alpha^2p^2][(p^{}_0+k^{}_0)^2+\alpha^2({\bm k}+{\bm p})^2]},\;\;\;\;\;\;
\end{eqnarray}
where $\alpha= Z^{-1}_1/Z^{-1}_0$. The integration can be performed with some effort, but we restrict our attempts to a more direct task, i.e. we  calculate only the modified optical conductivity $\tilde\sigma$ analogously to the optical conductivity of ideal graphene as we did before. We take the derivative with respect to $k$ under the integral and employ residue theorem in order to integrate out loop frequency $p^{}_0$. At the end of the calculation we arrive at
\begin{eqnarray}
\label{eq:CorrCond1}
\tilde\sigma = \tilde{e}^2 Z^{2}_0 \sigma^{}_0.
\end{eqnarray}
Surprisingly, apart from the vertex renormalization only renormalization of the frequency contributes to the modificated conductivity. Therefore our task reduces to the calculation of the renormalization factors $Z_0$ and $\tilde{e}$ which is performed perturbativelly below. 

\section{Calculation of renormalization factors}
\label{sec:Calc}

\begin{figure}[t]
\includegraphics[width=8.cm]{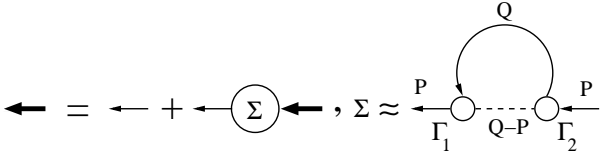}
\caption{Dressed fermionic Greens function and leading order diagram of the fermionic self-energy. Dashed lines denote the $\langle\Lambda\Lambda\rangle$-correlator and white circles the $\bar\Psi\Lambda\Psi$-vertices.}
\label{fig:SEF}
\end{figure}
\begin{figure}[t]
\includegraphics[height=2.2cm]{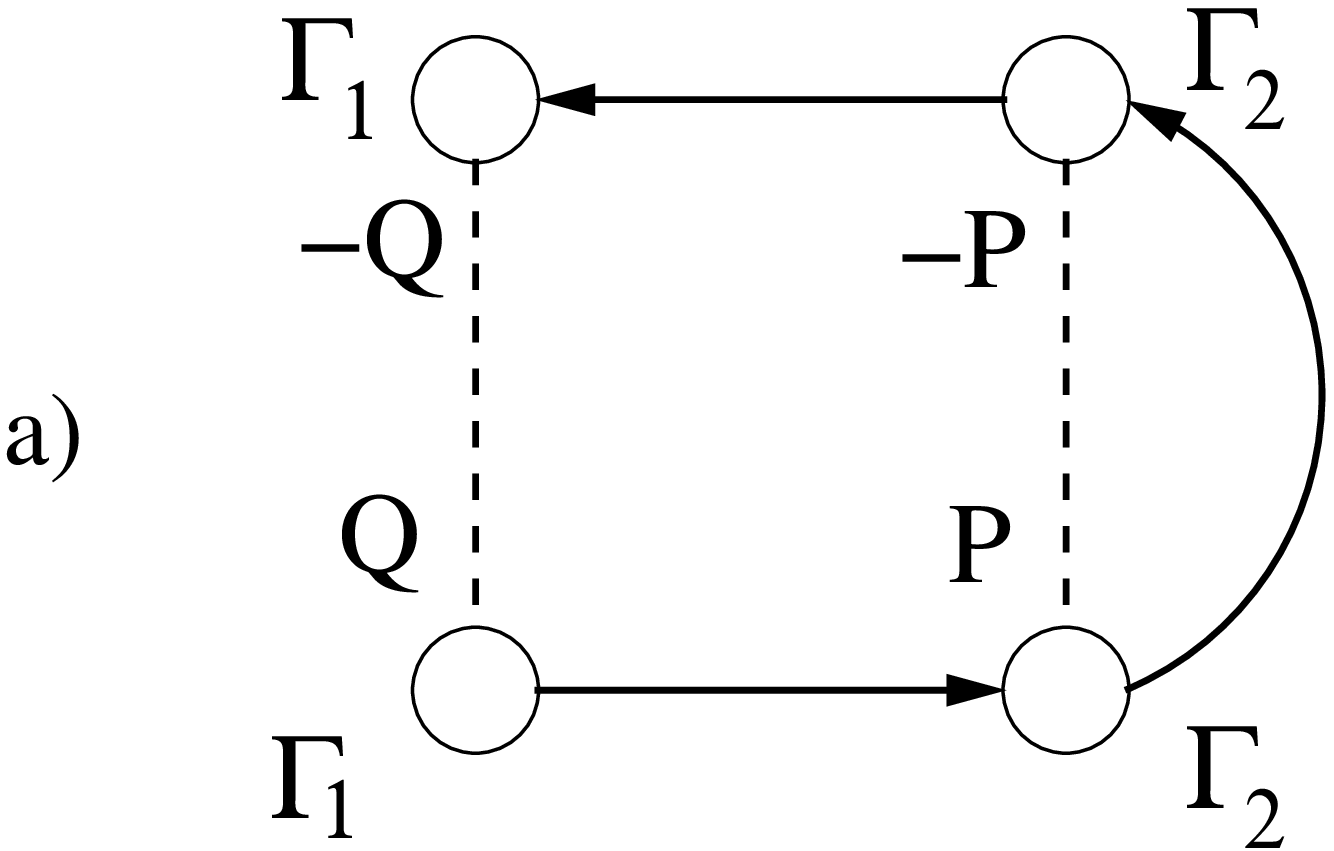}
\hspace{0.5 cm}
\includegraphics[height=2.2cm]{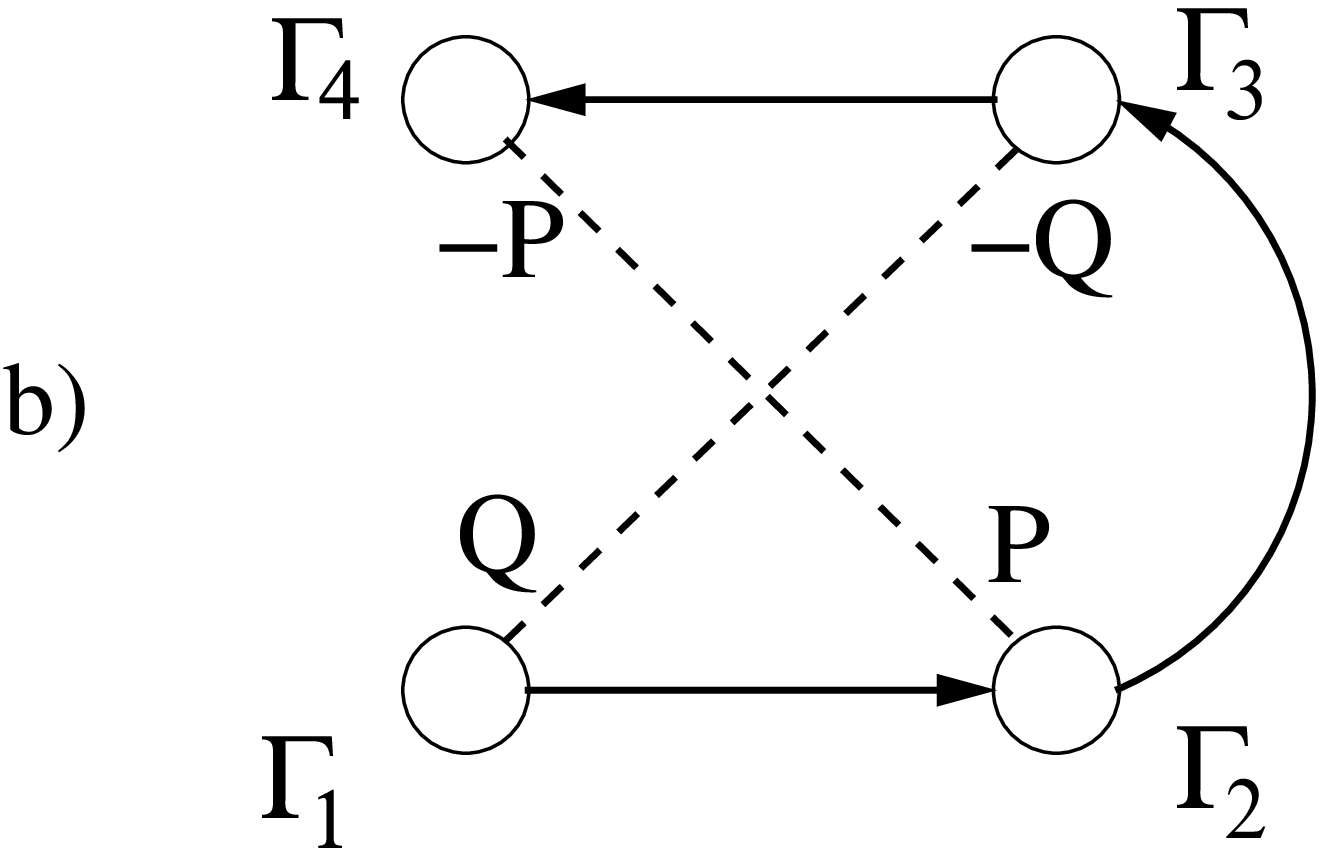}
\caption{Second order self-energy corrections at zero external momenta.}
\label{fig:SECII}
\end{figure}

In order to set up perturbative calculations we have to average over the lattice deformations. There are two possible ways to implement such averaging:  The replica trick and supersymmetry approach. The calculation below is based on the replica-trick approach.

According to the replica trick we introduce $N$ copies of fermions $\Psi^\alpha$, $\alpha=1,2,\dots N$ with the same action
\begin{eqnarray}
\nonumber
\bar{\cal S} &=& -\intop_{Q}~\bar\Psi^{\alpha}_{Q}(q^{}_0\gamma^{}_0+\gamma\cdot{\bm q})\Psi^{\alpha}_{Q} \\
\nonumber
&-&  \intop_{Q} \intop_{P}~\Lambda^{}_{P}\bar\Psi^{\alpha}_{{P}+{Q}} \Gamma(P+Q,P,Q) \Psi^{\alpha}_{Q}\\
\label{eq:ReplAct}
&-&\intop_{Q}\intop_{P}~A^{}_P \bar\Psi^{\alpha}_{P+Q}\gamma^{}_0\Psi^{\alpha}_{Q}.
\end{eqnarray}
Then we will calculate the diagrams describing renormalization of the fermionic propagator and electron-photon vertex function and perform limit $N\to0$ at the end of the calculation. As the result of this procedure all contributions containing factors proportional to any positive power of $N$ will vanish. These include for instance contributions arising from diagrams containing closed fermionic loops. 

\begin{figure}[t]
\includegraphics[width=6.cm]{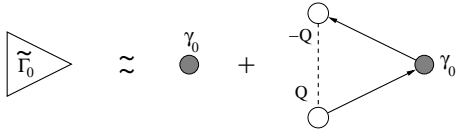}
\caption{Leading order perturbative contribution to the amputated electron-phonon coupling vertex. Dashed lines denote the $\langle\Lambda\Lambda\rangle$-correlator and white circles the $\bar\Psi\Lambda\Psi$-vertices.}
\label{fig:PhV}
\end{figure}

\begin{figure}[t]
\includegraphics[height=3cm]{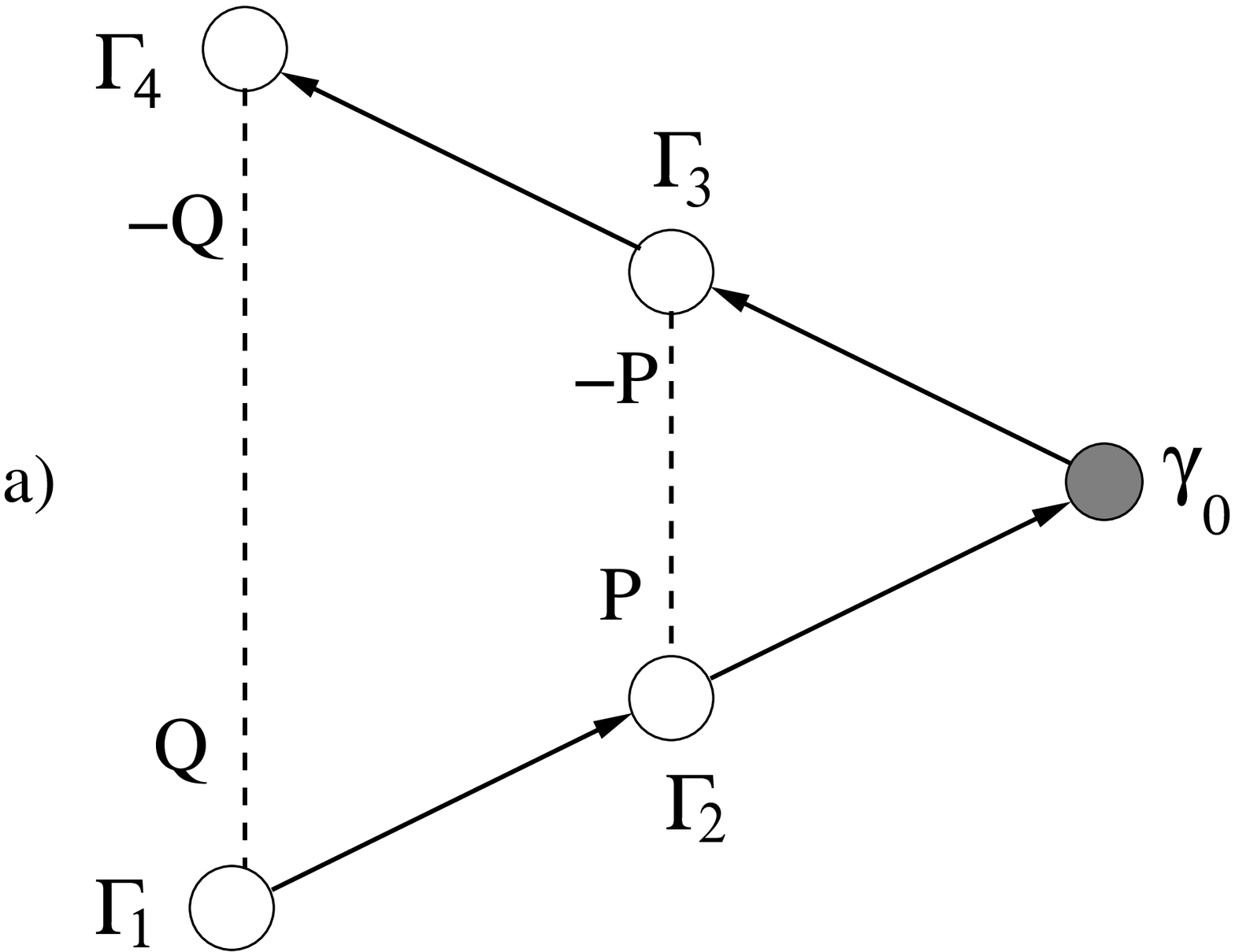}
\hspace{0.3cm}
\includegraphics[height=3cm]{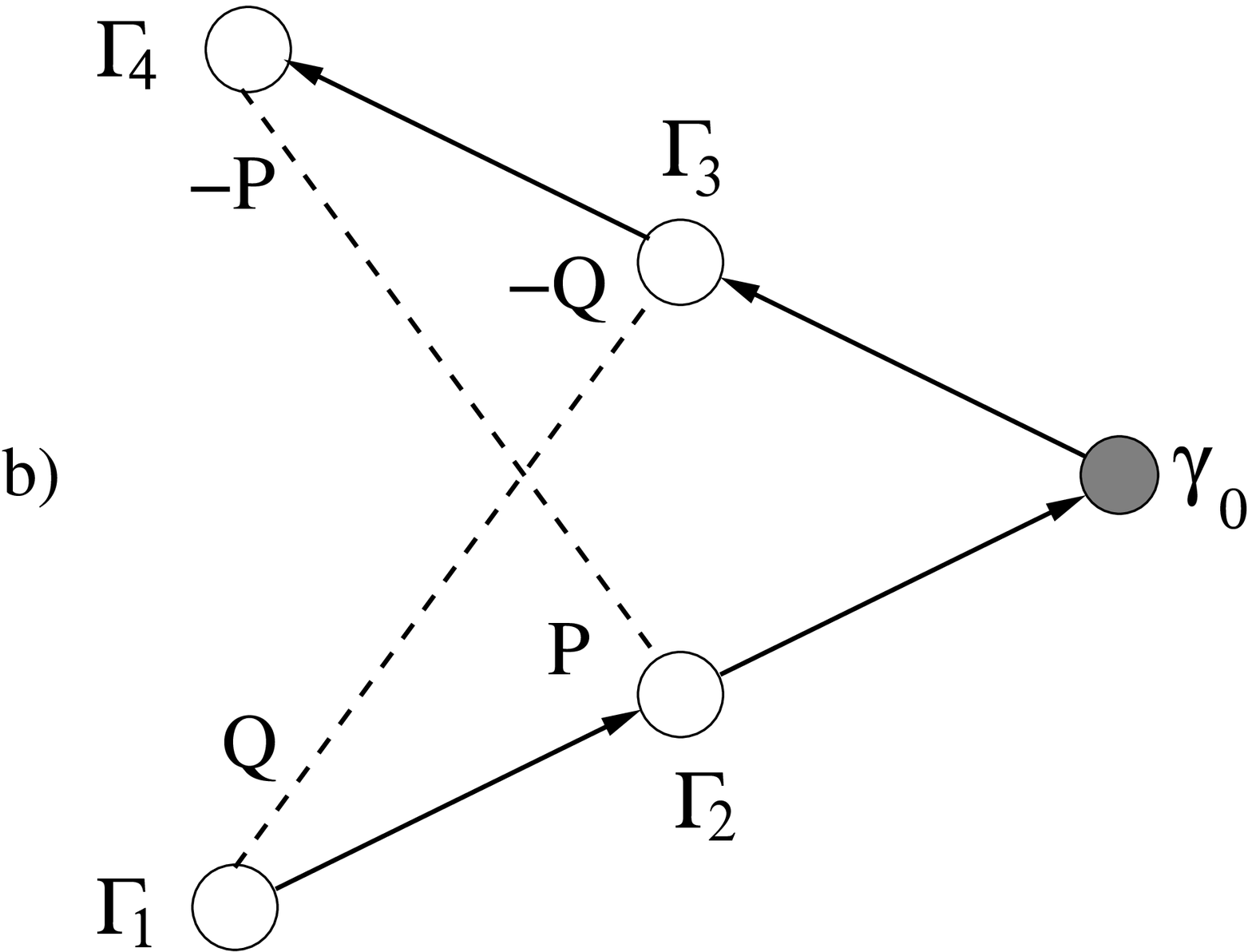}
\includegraphics[height=3cm]{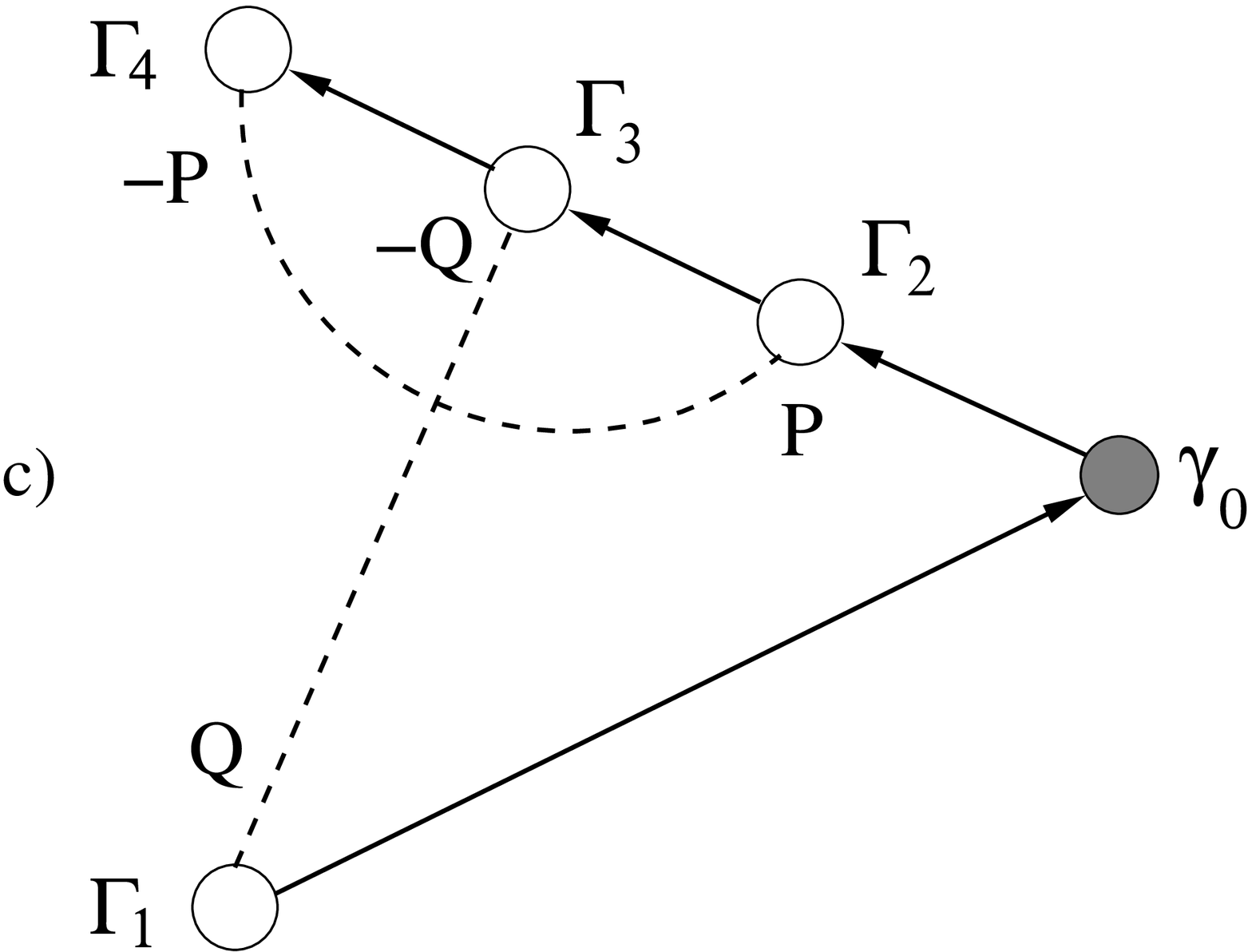}
\hspace{0.3cm}
\includegraphics[height=3cm]{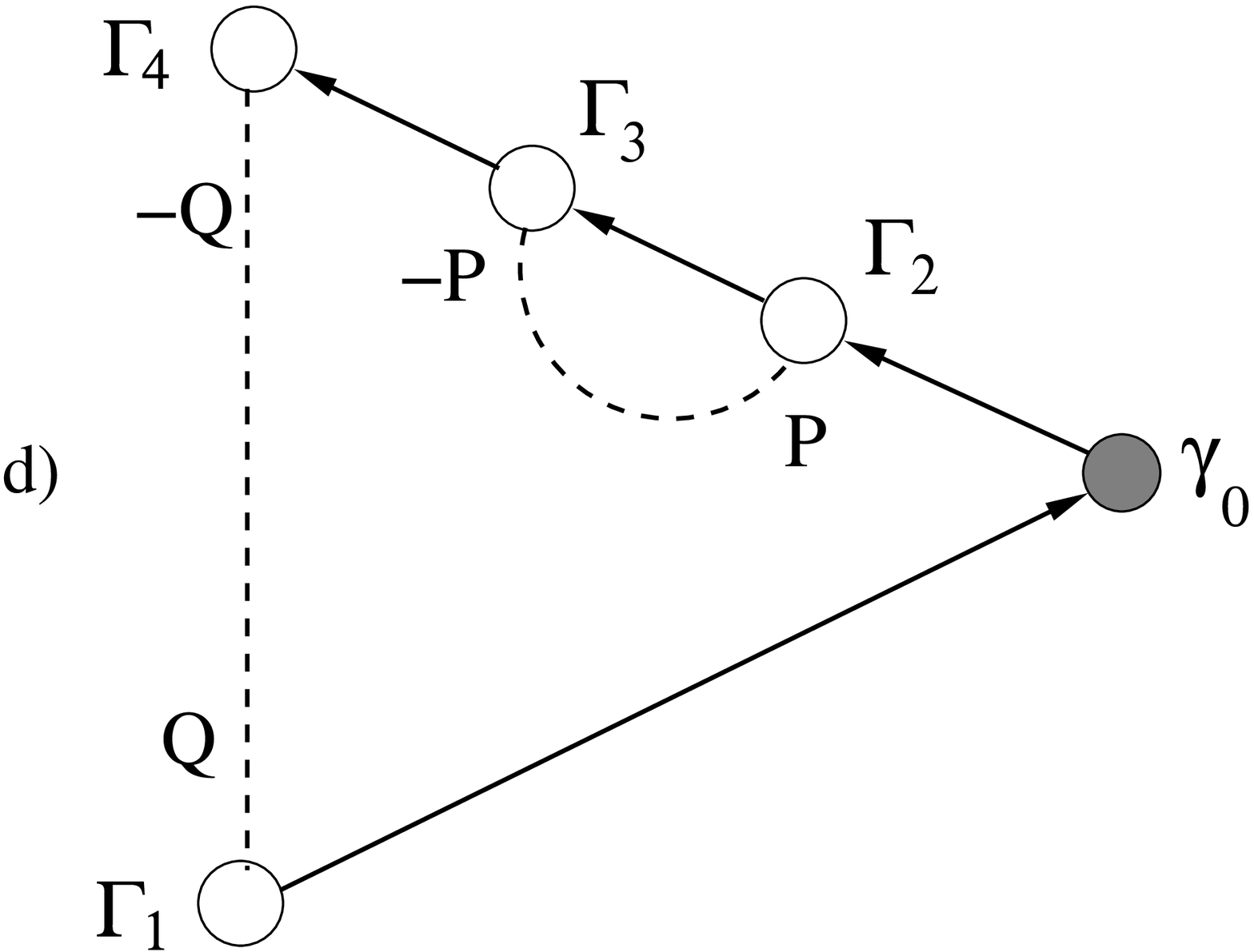}
\caption{Second order vertex correction diagrams at zero external momenta. The diagrams in the second row should be counted twice due to the mirror symmetry.}
\label{fig:VCII}
\end{figure}

The diagrams of the fermionic self-energy to the order 1 in replica indices and order $g^4_0$ in lattice deformations strength are shown in Figs.~\ref{fig:SEF} and \ref{fig:SECII}. Correspondingly, the same order diagrams of vertex corrections are depicted in Figs.~\ref{fig:PhV} and~\ref{fig:VCII}. Retaining only frequency dependence (${\bm p}=0$) in the analytical expressions for this contributions we obtain for the leading self-energy contribution (Fig.~\ref{fig:SEF})
\begin{eqnarray}
\label{eq:SEF}
\Sigma^{(2)}(p^{}_0)=\frac{g^2_0}{4}\intop_{\bm q} ~\Gamma^{}_1G^{}_0(p^{}_0,{\bm q})\Gamma^{}_2 F({\bm q}),
\end{eqnarray}
where $\Gamma^{}_1=\Gamma^{}_2=\gamma\cdot{\bm q}$ and
\begin{eqnarray}
F({\bm q}) = \frac{1}{q^2+\mu^2},
\end{eqnarray}
denotes the momentum dependent part of the $\langle\Lambda\Lambda\rangle-$correlator defined in Eq.~(\ref{eq:LLcorr}). The diagram of the next order in $g^{}_0$ depicted in Fig.~{\ref{fig:SECII}a} reads
\begin{eqnarray}
\nonumber
\Sigma^{(2)}_{1}(p^{}_0) &=& \frac{g^4_0}{16}\intop_{\bm q}~F({\bm q})\intop_{\bm p}~F({\bm p})~\Gamma^{}_1G^{}_0(p^{}_0,{\bm q})\Gamma^{}_2\\
\nonumber
&&\times G^{}_0(p^{}_0,{\bm q}+{\bm p})\Gamma^{}_2G^{}_0(p^{}_0,{\bm q})\Gamma^{}_1,
\end{eqnarray}
where $\Gamma^{}_1=\gamma\cdot{\bm q}$ and $\Gamma^{}_2=\gamma\cdot(2{\bm q}+{\bm p})$, while the diagram shown in Fig.~{\ref{fig:SECII}b} writes
\begin{eqnarray}
\nonumber
\Sigma^{(2)}_{2}(p^{}_0) &=& \frac{g^4_0}{16}\intop_{\bm q}~F({\bm q})\intop_{\bm p}~F({\bm p})~\Gamma^{}_1G^{}_0(p^{}_0,{\bm q})\Gamma^{}_2\\
\nonumber
&&\times G^{}_0(p^{}_0,{\bm q}+{\bm p})\Gamma^{}_3G^{}_0(p^{}_0,{\bm p})\Gamma^{}_4,
\end{eqnarray}
with the vertices $\Gamma^{}_1=\gamma\cdot{\bm q}$, $\Gamma^{}_2=\gamma\cdot(2{\bm q}+{\bm p})$, $\Gamma^{}_3=\gamma\cdot(2{\bm p}+{\bm q})$ and $\Gamma^{}_4 = \gamma\cdot{\bm p}$. Eventually we obtain for the contributions to the wave-function renormalization factor 
\begin{eqnarray}
\left.\frac{\partial}{\partial p^{}_0}\Sigma^{(2)}(p^{}_0)\right|_{p^{}_0=0}&=& -\hat{e} \\
\left.\frac{\partial}{\partial p^{}_0}\Sigma^{(4)}_{1}(p^{}_0)\right|_{p^{}_0=0}&=&-9{\hat e}^2,\\
\left.\frac{\partial}{\partial p^{}_0}\Sigma^{(4)}_{2}(p^{}_0)\right|_{p^{}_0=0}&=&-\frac{15}{2}{\hat e}^2,
\end{eqnarray}
where we define 
\begin{equation}
\label{eq:ElChargeRen}
{\hat e} = \frac{g^2_0}{8\pi}\log\frac{\lambda}{\mu},
\end{equation}
with $\lambda$ denoting some upper momentum cutoff. Therefore we obtain for the wave-function renormalization to the second order in $g^2_0$ 
\begin{equation}
\label{eq:CorrZ02}
Z^{-1}_0 \approx 1 + {\hat e} + \frac{33}{2}{\hat e}^2+{\cal O}({\hat e}^3).
\end{equation}

Now we look at the renormalization of the electron-photon vertex function. Diagrammatically the leading order correction is given by the second term on the right hand side of the diagram shown in Fig.~\ref{fig:PhV}. According to Eq.~(\ref{eq:CorrCond1}) the main corrections to the conductivity arise from the momentum independent part of the vertex function. We obtain for the  vertices $\Gamma^{}_1 = \Gamma^{}_2 = \gamma\cdot{\bm q}.$ The algebraic expression for the diagram depicted in Fig.~\ref{fig:PhV} is given by
\begin{equation}
\label{eq:VCeq1}
\tilde\Gamma^{(1)}(0) = \frac{g^2_0}{4}\intop_{\bm q}~\Gamma^{}_1 G^{}_0({\bm q}) \Gamma^{}_0 G^{}_0({\bm q}) \Gamma^{}_2F({\bm q}),
\end{equation}
which yields after the evaluation 
\begin{equation}
\label{eq:VertCorrection5}
\tilde\Gamma^{(1)}(0) = {\hat e} \gamma^{}_0,
\end{equation}
where ${\hat e}$ from Eq.~(\ref{eq:ElChargeRen}) is introduced and acquires the meaning of the leading order elementary charge renormalization.


Second order vertex corrections can be calculated from the diagrams shown in Fig.~\ref{fig:VCII}. Due to the mirror symmetry diagrams c) and d) depicted in the second row in Fig.~\ref{fig:VCII} should be counted twice. Let us first consider the contribution that arises from the diagram depicted in Fig.~{\ref{fig:VCII}a} with two parallel ladder rungs. At zero external momenta we obtain for the vertices $\Gamma^{}_1=\Gamma^{}_4=\gamma\cdot{\bm q}$ and $\Gamma^{}_2=\Gamma^{}_3=\gamma\cdot(2{\bm q}+{\bm p})$. 
We obtain for the correction
\begin{eqnarray}
\nonumber
\tilde\Gamma^{(2)}_{1}(0) &=& \frac{g^4_0}{16}\intop_{\bm q}~F({\bm q})\intop_{\bm p}~F({\bm p})~\Gamma^{}_1 G^{}_0({\bm q}) \Gamma^{}_2 G^{}_0({\bm p}+{\bm q})\\
\label{eq:SecOrdVert1}
&&\times \Gamma^{}_0  G^{}_0({\bm p}+{\bm q}) \Gamma^{}_2 G^{}_0({\bm q}) \Gamma^{}_1,
\end{eqnarray}
which, after performing integrations, yields
\begin{equation}
\label{eq:SecOrdVert2}
\tilde\Gamma^{(2)}_{1}(0) = 2{\hat e}^2\gamma^{}_0.
\end{equation}

For diagrams depicted in Figs.~\ref{fig:VCII}b,~c and~d we proceed similarly. In the case of diagram b) we have the following expressions for the vertices: $\Gamma^{}_1 = \gamma\cdot{\bm q}$, $\Gamma^{}_2 = \gamma\cdot(2{\bm q}+{\bm p})$, $\Gamma^{}_3 = \gamma\cdot(2{\bm p}+{\bm q})$ and $\Gamma^{}_4 = \gamma\cdot{\bm p}$. Therefore the expression for this correction reads
\begin{eqnarray}
\nonumber
\tilde\Gamma^{(2)}_{2}(0) &=& \frac{g^4_0}{16}\intop_{\bm q}~F({\bm q})\intop_{\bm p}~F({\bm p})~\Gamma^{}_1 G^{}_0({\bm q}) \Gamma^{}_2 G^{}_0({\bm p}+{\bm q})\\
\label{eq:SecOrdVert4}
&&\times\Gamma^{}_0  G^{}_0({\bm p}+{\bm q}) \Gamma^{}_3 G^{}_0({\bm p}) \Gamma^{}_4,
\end{eqnarray}
with the result
\begin{equation}
\label{eq:SecOrdVert5}
\tilde\Gamma^{(2)}_{2}(0) = \frac{5}{2}{\hat e}^2\gamma^{}_0.
\end{equation}
For diagram c) we have the following vertices: $\Gamma^{}_1=\gamma\cdot{\bm q}$, $\Gamma^{}_2=\gamma\cdot(2{\bm q}+{\bm p})$, $\Gamma^{}_3 = \gamma\cdot(2{\bm p}+{\bm q})$ and $\Gamma^{}_4=\gamma\cdot{\bm p}$, whereas the expression for the correction reads
\begin{eqnarray}
\nonumber
\tilde\Gamma^{(2)}_{3}(0) &=& \frac{g^4_0}{16}\intop_{\bm q}~F({\bm q})\intop_{\bm p}~F({\bm p})~\Gamma^{}_1 G^{}_0({\bm q}) \Gamma^{}_0 G^{}_0({\bm q}) \Gamma^{}_2\\
\label{eq:SecOrdVert6}
&&\times G^{}_0({\bm p}+{\bm q}) \Gamma^{}_3 G^{}_0({\bm p}) \Gamma^{}_4,
\end{eqnarray}
which yields the result
\begin{equation}
\label{eq:SecOrdVert7}
\tilde\Gamma^{(2)}_{3}(0) = \frac{5}{2}{\hat e}^2\gamma^{}_0.
\end{equation}

Finally, diagram d) from Fig.~\ref{fig:VCII} can be written algebraically as follows:
\begin{eqnarray}
\nonumber
\tilde\Gamma^{(2)}_{4}(0) &=& \frac{g^4_0}{16}\intop_{\bm q}~F({\bm q})\intop_{\bm p}~F({\bm p})~\Gamma^{}_1 G^{}_0({\bm q}) \Gamma^{}_0 G^{}_0({\bm q}) \Gamma^{}_2\\
\label{eq:SecOrdVert8}
&&\times G^{}_0({\bm p}+{\bm q}) \Gamma^{}_3 G^{}_0({\bm q}) \Gamma^{}_4,
\end{eqnarray}
where the vertices $\Gamma^{}_{1\cdots4}$ are given by $\Gamma^{}_1=\Gamma^{}_4=\gamma\cdot{\bm q}$ and $\Gamma^{}_2=\Gamma^{}_3=\gamma\cdot(2{\bm q}+{\bm p})$. Evaluation of Eq.~(\ref{eq:SecOrdVert8}) yields
\begin{equation}
\label{eq:SecOrdVert9}
\tilde\Gamma^{(2)}_{4}(0) = \frac{7}{2}{\hat e}^2\gamma^{}_0.
\end{equation}

Hence, the second order contribution to the vertex function becomes
\begin{equation}
\tilde\Gamma^{(2)}(0) = \tilde\Gamma^{(2)}_{1}(0)+\tilde\Gamma^{(2)}_{2}(0)+2\tilde\Gamma^{(2)}_{3}(0)+2\tilde\Gamma^{(2)}_{4}(0),
\end{equation}
such that the dressed vertex function can be written as a series in $\hat e$:
\begin{equation}
\label{eq:CorrVer1}
\tilde\Gamma^{}_{0} = \tilde{e}\gamma^{}_0 \approx e\left(1+{\hat e}+\frac{33}{2}{\hat e}^{2}+{\cal O}({\hat e}^3)\right)\gamma^{}_0.
\end{equation}
This expression reproduces exactly the result which we have obtained for the dressed vertex function in Eq.~(\ref{eq:CorrZ02}). Therefore we obtain from Eqs.~(\ref{eq:CorrCond1}), (\ref{eq:CorrVer1}) and (\ref{eq:CorrZ02}) for the modified conductivity
\begin{equation}
\tilde\sigma \approx \left(1+{\cal O}({\hat e}^3)\right)\sigma^{}_0,
\end{equation}
i.e. the leading correction is of the order $g^6_0$. However, we can show to every order in perturbative expansion that corrections arising from the propagator renormalization are exactly canceled by their counterparts departing from the electron-photon vertex renormalization. Consider the definition of the quasi-particle weight $Z^{-1}_0$ given in Eq.~(\ref{eq:WaveFuncRen}). 
Since the lattice deformations are static, all propagators inside the diagram depend only on the external Matsubara-frequency which thus becomes an independent parameter. Hence the derivative with respect to the Matsubara frequency should be applied to every propagator. Taking such a derivative of an average free propagator 
\begin{equation}
\langle G^{}_0(q^{}_0,{\bm q}+{\bm k})\rangle^{}_{\bm k} = \intop_{\bm k}\frac{q^{}_0\gamma^{}_0+(q^{}_\mu+k^{}_\mu)\gamma^{}_\mu}{q^2_0+({\bm q}+{\bm k})^2},
\end{equation}
at zero external momentum and frequency we obtain
\begin{eqnarray}
\label{eq:ExactRel}
\left.\frac{\partial}{\partial q^{}_0} \langle G^{}_0(q^{}_0,{\bm q}+{\bm k})\rangle^{}_{\bm k}\right|_{Q=0} = 
-\langle G^{}_0(0,{\bm k})\gamma^{}_0 G^{}_0(0,{\bm k})\rangle^{}_{\bm k}.
\;\;
\end{eqnarray}
Equation~(\ref{eq:ExactRel}) suggests that the expressions under the integrals must be equal up to an irrelevant constant. Therefore each derivative of the free propagator with respect to the external frequency  generates upon sending external momenta and frequency to zero a bare electron-photon vertex. An irreducible $n-$th order diagram of the electronic self-energy contains $2n-1$ electronic propagators. Therefore by applying a derivative with respect to the frequency $q^{}_0$ to such a diagram, $2n-1$ irreducible corrections to the electron-photon interaction vertex are generated. This mimics term by term a perturbative series for the dressed vertex. This can be seen very clearly if we look at self-energy diagrams depicted in Fig.~\ref{fig:SECII}. Replacing electronic propagators successively by a bare electron-phonon interaction vertex and putting external momenta to zero we reproduce exactly vertex-correction diagrams shown in Fig.~\ref{fig:VCII}. We therefore can link each $n$-th ($n\geqslant1$) term in the perturbative series of self-energy to the corresponding vertex function correction:
\begin{equation}
\left.\frac{\partial}{\partial q^{}_0}\Sigma^{(n)}(Q)\right|_{Q=0} = -\tilde\Gamma^{(n)}_0(0,0,0).
\end{equation}
Summing over all $n$ and subtracting $\gamma^{}_0$ on both sides we then can assemble all contributions arriving at
\begin{equation}
\label{eq:FinRres}
\left.\frac{\partial}{\partial q^{}_0}G^{-1}(Q)\right|_{Q=0} = \tilde\Gamma^{}_0(0,0,0).
\end{equation}
On the right-hand side we have the charge renormalization $\tilde e\gamma^{}_0$ defined in Eq.~(\ref{eq:DresVert}), while the left-hand side represents the wave-function renormalization factor $Z^{-1}_0\gamma^{}_0$ due to Eq.~(\ref{eq:DresProp}). Therefore Eq.~(\ref{eq:FinRres}) postulates the equality 
$$
Z^{-1}_0 = \tilde{e},
$$
which leads to the exact result for the modified conductivity:
\begin{equation}
\tilde{\sigma} = \sigma^{}_0. 
\end{equation}
Importantly Eq.~(\ref{eq:FinRres}) is obtained without special emphasize on a disorder type and is not restricted to the considered type. The only requirement we need is that the corresponding term should not violate the chiral symmetry of the pure graphene Hamiltonian and must be a quenched disorder.

\section{Conclusions}

In the present paper we address the question of the effect which random deformations may have on the transport in graphene. 
The common believe is that surface corrugations in graphene influence its electronic transport properties, mainly the optical conductivity. 
It is possible to describe deformations in graphene by a gauge field that couples to the fermions living on the two dimensional sheet. We have performed perturbative calculations of the corrections due to lattice deformations to the optical conductivity. Our results contrast the suggestions made in Refs.~[\onlinecite{Cortijo2009,VozKatsGuin2010}] where a substantional effect of the defects on the conductivity is proposed. We have found that the minimal conductivity is robust with respect to the surface corrugations. 

\section*{ACKNOWLEDGEMENTS}

We acknowledge financial support by the DPG-grant ZI 305/5-1.

\appendix
\section*{Appendix}
Below be evaluate the irreducible polarization of ideal graphene starting with Eq.~(\ref{eq:IrPol}).
Upon performing the trace over the pseudo-spin space we arrive at
\begin{eqnarray}
\nonumber
\Pi(K) = 2\intop_P~\frac{2p^{}_0(p^{}_0+k^{}_0)-P\cdot(P+K)}{P^2(P+K)^2},
\end{eqnarray}
where $P\cdot K=p^{}_0k^{}_0+{\bm p}\cdot{\bm k}$. Employing the Feynman parametrization
$$
\frac{1}{AB}=\intop_0^1~\frac{dx}{[xA+(1-x)B]^2},
$$
and shifting $P\to P-x K$ we symmetrize the denominator with respect to $P$. Therefore odd powers of $P$ appearing in the numerator may be dropped. We arrive at
\begin{equation}
\nonumber
\Pi(K) = 2\intop_P~\intop_0^1 dx~\frac{2p^2_0 - P^2+x(1-x)[K^2-2k^2_0]}{[P^2+x(1-x)K^2]^2}.
\end{equation}
Exploiting the rotational invariance we replace $p^{2}_0 = P^2/3$ and use formulas of dimensional regularization:
\begin{eqnarray}
\nonumber
\intop\frac{d^dk}{(2\pi)^d} \frac{1}{(k^2+\Delta)^n} &=& \frac{1}{(4\pi)^{d/2}}
\frac{\Gamma(n-\frac{d}{2})}{\Gamma(n)\Delta^{n-{d/2}}},\\
\nonumber
\intop\frac{d^dk}{(2\pi)^d} \frac{k^2}{(k^2+\Delta)^n} &=& \frac{1}{(4\pi)^{d/2}}\frac{d}{2} \frac{\Gamma(n-\frac{d}{2}-1)}{\Gamma(n)\Delta^{n-{d/2}-1}},\;\;
\end{eqnarray}
which yields after integrating out $x$ the result of Eq.~(\ref{eq:IrPolRes}).

\end{document}